\documentclass[12pt]{article}
\usepackage[cp1251]{inputenc}
\usepackage[english]{babel}
\usepackage{graphicx,graphics}
 0
\begin{document}

\title{
{\sf \bf Generalized Zero Range Potentials and Multi-Channel
Electron-Molecule Scattering}}

\author{  S.B. Leble
\small \\ Theoretical Physics and Mathematical Methods Department,
\small\\ Technical University of Gdansk, ul, Narutowicza 11/12, Gdansk,  Poland,
\small \\ leble@mifgate.pg.gda.pl \\  \\[2ex]
S. Yalunin \small\\ Theoretical Physics Department, \small\\
Kaliningrad State University, A. Nevsky st. 14, Kaliningrad,
Russia, \small \\ yalunin@bk.ru}
 \maketitle

\renewcommand{\abstractname}{\small Abstract}
\begin{abstract}
A multi-channel scattering problem is studied from a point of view
of integral  equations system. The system appears while natural
one-particle wave function equation
 of the electron under action of a potential with non-intersecting ranges is considered.
 Spherical functions basis expansion of the potentials introduces partial amplitudes
 and corresponding radial functions. The approach is generalized to multi-channel case
 by a matrix formulation in which a state vector component is associated with
a scattering channel.
 The  zero-range potentials naturally enter the scheme when the class of operators of
multiplication is widen to distributions.
Oscillations and rotations are incorporated into the scheme.
\end{abstract}

\thispagestyle{empty}
\section{Introduction}
In 1936, Fermi $\cite{Fermi}$ proposed zero range potential (ZRP)
model to study neutron scattering in hydrogen-containing
substances. Since then, ZRP approach have been developed to widen
limits of this pioneering treatment (for a review see Demkov and
Ostrovsky $\cite{DO1975}$, Drukarev $\cite{D1978}$, Albeverio et
al. $\cite{AlGa1988}$). The advantage of the theory is the
possibility of obtaining an exact solution of scattering problem.
Recently, Baltenkov $\cite{Balt}$ have generalized the ZRP method
for the case of non-zero orbital moments (see \cite{BW} )including
combinations of potentials . De Prunel$\acute{\hbox{e}}$
$\cite{Prunele}$ have proposed other solvable non-zero range
potentials, which involve higher partial waves.

The aim of this paper relates to the other limitation of the real
scattering phenomenon. Following the main ideas of the ZRP method
we introduce a matrix ZRP potential, (Section $\ref{mZRP}$), for a
multichannel problem (see, for instance, Lane $\cite{La1980}$).
The matrix ZRP is conventionally represented as the boundary
condition on the matrix wavefunction at some point. Presented
matrix potential generalizes ordinary matrix ZRP, which was
proposed by Demkov and Ostrovsky $\cite{DO1970}$ (see also
$\cite{LY}$), for the case of the {\it s,p,d,} etc. target states.
We also present simple method of deriving scattering amplitude for
a multi-center problem (Section $\ref{mcp}$) and consider two
matrix ZRPs problem (Section $\ref{TZRP}$). As an important
example we consider applications to a diatomic molecule. The
formulae for differential and integral cross sections of the
electron-vibrational excitations are summarized in the Section
$\ref{DICS}$. In the Section $\ref{Apl}$ we present the results of
our numerical calculations  for the molecule $\hbox{H}_2$ as well
as plots for electron-vibrational cross-sections.

\section{Nonoverlaping potentials and multi-center problem}\label{mcp}
\subsection{One-center case}
Let us consider the scattering problem for matrix wavefunction
$y({\bf r},K{\bf n}_0)$ and short-range matrix operator,
which is $U$ at $r\leq d$ and zero at
$r> d$. The atomic units are used throughout the
present paper.
In the interior (i.e. at $r<d$)
\begin{equation} \label{Eqy}
y({\bf r},K{\bf n}_0)=e^{iK{\bf n}_0\cdot{\bf r}}-
(2\pi)^{-1} \int
\frac{e^{iK |{\bf r}-{\bf r}'|}}{\scriptstyle |{\bf r}-{\bf r}'|}
U y({\bf r}',K{\bf n}_0) d{\bf r}',
\end{equation}
where $K=\hbox{diag}(k_{n})$, $k_n=\sqrt{k_0^2-2E_n}$ are electron
momenta in the channels $n$, $E_n$ are excitation energies, and
${\bf n}_0$ indicates incoming electron direction. The matrix
wavefunction $y({\bf r},K{\bf n}_0)$ in the exterior region (i.e.
at $r>d$) is a solution of the Helmholtz equation and is
determined by smooth matching condition at the boundary (i.e. at
$r=d$).

In order to construct solutions in the exterior region we
introduce a matrix function $H({\bf r},{\bf u};K)$ as a solution
of the equation
\begin{equation} \label{eqH}
\frac{e^{iK|{\bf r}-{\bf r}'|}}
{\scriptstyle|{\bf r}-{\bf r}'|}=
\int_{\Omega} H({\bf r},{\bf u};K) e^{-iK{\bf u}\cdot{\bf r}'}
d\Omega_{\bf u},\hspace{10mm}r'<r,
\end{equation}
where ${\bf u}$ is unit vector. This diagonal matrix function is
invariant under rotation transformations of the vectors ${\bf r},
{\bf u}$ and has the following asymptotic behavior at infinity
\begin{equation}
H({\bf r},{\bf u};K) \stackrel{\rm r \rightarrow \infty}{\sim}
\frac{e^{iKr}}{r}
\, \delta({\bf n}-{\bf u}),
\end{equation}
here ${\bf n}=r^{-1}{\bf r}$, and $\delta({\bf n})$ denotes
delta function of the angle variable ${\bf n}$.
Solution in the exterior region, i.e.
$y({\bf r},K{\bf n}_0)$ at $r>d$, can be constructed in terms of
the matrix function $H({\bf r},{\bf u};K)$

\begin{equation} \label{EQy}
y({\bf r},K{\bf n}_0)=e^{iK{\bf n}_0\cdot {\bf r}}+
\int_{\Omega}
 H({\bf r},{\bf u};K) F({\bf u},{\bf n}_0;K) d\Omega_{\bf u},
\end{equation}
where matrix amplitude $F({\bf n},{\bf n}_0;K)$
is given by the expression
\begin{equation} \label{EQF}
F({\bf n},{\bf n}_0;K)=-(2\pi)^{-1}
\int e^{-iK{\bf n}\cdot {\bf r}}
U y({\bf r},K{\bf n}_0) d{\bf r},
\end{equation}
in which ${\bf n}$ is outgoing electron direction.

\subsection{Multi-center case}
Consider $N$ nonoverlapping short-range matrix operators $
e^{-i{\bf p}\cdot{\bf R}_i}\ U_i\ e^{i{\bf p}\cdot{\bf R}_i}, $
here ${\bf p}$ describes the momentum operator $-i\nabla$. Suppose
that the interaction regions are limited by spheres of the
radiuses $d_i$ with centers at the points ${\bf R}_i$. Requirement
of nonoverlapping is written down as $d_i+d_j \leq |{\bf R}_i-{\bf
R}_j|$. Denote the matrix wavefunction in the region $|{\bf
r}-{\bf R}_i|<d_i$ by the expression $e^{-i{\bf p}\cdot {\bf R}_i}
\Psi_i({\bf r})$. The matrix wavefunctions $\Psi_i({\bf r})$
satisfy the following equations in the region $r<d_i$
$$
\Psi_i({\bf r})=e^{iK {\bf n}_0 \cdot ({\bf r}+{\bf R}_i)}-
(2\pi)^{-1}\ \sum_{j=1}^{N}
\int
\frac{e^{iK|{\bf r}-{\bf r}'+{\bf R}_i-{\bf R}_j|}}
{\scriptstyle |{\bf r}-{\bf r}'+{\bf R}_i-{\bf R}_j|}\,
U_j \Psi_j({\bf r}')\, d{\bf r}',
$$
where integration must be performed over $r' < d_j$. Taking into
account the  condition of nonoverlapping and the  equation
($\ref{eqH}$) we obtain the expressions
$$
\Psi_i({\bf r})=e^{iK{\bf n}_0\cdot ({\bf r}+{\bf R}_i)}-
(2\pi)^{-1} \int
\frac{e^{iK |{\bf r}-{\bf r}'|}}{\scriptstyle |{\bf r}-{\bf r}'|}
U_i \Psi_i({\bf r}') d{\bf r}' -
$$
$$
(2\pi)^{-1} \sum_{j \neq i}^N
\int \int_{\Omega} H({\bf R}_i-{\bf R}_j,{\bf u};K)
e^{iK{\bf u}\cdot ({\bf r}-{\bf r}')}U_j\Psi_{j}({\bf r}') d{\bf r}'
d\Omega_{\bf u}.
$$
To express the matrix wavefunctions $\Psi_i({\bf r})$, let us use
solutions of the correspondent one-center problems. In terms of
the matrices $y_i({\bf r},K{\bf u})$, the  matrices $\Psi_i({\bf
r})$ become
$$
\Psi_i({\bf r})=\int_{\Omega} \int_{\Omega} y_i({\bf r},K{\bf n})
F_i^{-1}({\bf n},{\bf n}';K)\,
C_i({\bf n}',{\bf n}_0) d\Omega_{\bf n} d\Omega_{{\bf n}'},
$$
where $C_i({\bf n}',{\bf n}_0)$ are some constant matrices. Taking
into account Eqs. $(\ref{EQy})$ and $(\ref{EQF})$ we obtain
\begin{equation} \label{EQC}
\int_{\Omega} F_{i}^{-1}({\bf n},{\bf n}';K)\,
C_i({\bf n}',{\bf n}_0) d\Omega_{{\bf n}'}=
e^{iK {\bf n}_0\cdot {\bf R}_i} \delta({\bf n}-{\bf n}_0)+
\sum_{j \neq i}^N H({\bf R}_i -{\bf R}_j,{\bf n};K)
C_j({\bf n},{\bf n}_0).
\end{equation}
In order to calculate multi-center $F$-matrix for the
nonoverlapping potentials we only need to solve this matrix
equation in $N$ unknown matrices $C_i({\bf n},{\bf n}_0)$. The
multi-center $F$-matrix is given by
\begin{equation} \label{EQFmc}
F({\bf n},{\bf n}_0;K)=\sum_{i=1}^{N}
e^{-iK{\bf n}\cdot {\bf R}_i} C_{i}({\bf n},{\bf n}_0).
\end{equation}
We will omit from now the argument $K$ of the matrix $F$-function.

If we take into account only $s$-wave point interaction then
Eqs. $(\ref{EQC}),(\ref{EQFmc})$ are reduced to usual equations of
the ZRP theory.
Another solvable model was recently proposed by
de Prunel$\acute{\hbox{e}}$ $\cite{Prunele}$. In this model
nonoverlapping separable interactions are
localized on the spheres of the radiuses $d_i$ with centers at
the points ${\bf R}_i$. It is clear that the most essential aspects
of this model are obtained also by using the Eq. $(\ref{EQC})$.
Thus, presented formalism can be a basis of a
new solvable models creation.

\section{Matrix ZRP and boundary condition}\label{mZRP}

Now we examine closely the electron-center interaction on the
multi-channel level of approximation. If interaction preserves a
total angular momentum of electron-center system then the matrix
partial waves are also eigenfunctions of the total and projection
angular momentum operators. Further we consider for simplicity  a
case of the zero total angular momentum. Assume $l_n, m_n$ denote
angular momentum quantum numbers of isolated center in the channel
$n$. The isolated center states may be characterized as $s,p,d,$
etc. for $l_n=0,1,2,$ etc. respectively. Hence matrix function
$y({\bf r},K{\bf n}_0)\sim A({\bf n})$, where $A({\bf n})$ is the
diagonal matrix
\begin{equation} \label{EqTAM}
A({\bf n})=\hbox{diag} \left(
Y_{l_{n} m_{n}}({\bf n})\right).
\end{equation}
Thus, $l_n$ is also orbital momentum of the incident electron in
the channel $n$; $m_n$ is its projection on the axis $z$.

Following the main idea of the ZRP method, we suppose that matrix
interaction $U$ is localized at the center point (the point can be
considered also as a sphere of radius $d \rightarrow
0{\scriptstyle ^+}$). Therefore the matrix $y({\bf r},K{\bf n}_0)$
is a superposition of the regular and irregular Helmholtz
solutions. The asymptotic behavior at zero is given by (further we
use for simplicity the notations like
$\hbox{diag}(r^{x_n})=r^{\hbox{diag}(x_n)}$ and
$\hbox{diag}(y_n^{x_n})=\hbox{diag}(y_n)^{\hbox{diag}(x_n)}$.)
\begin{equation} \label{EqPsi}
y({\bf r},K{\bf n}_0)\stackrel{\rm r \rightarrow 0}{\sim}
A({\bf n})\left( (2L-E)!!\, r^{-L-E} -
\frac{r^{L}}{(2L+E)!!} W \right) A^+({\bf n}_0),
\end{equation}
where $E$ is unit matrix, $L=\hbox{diag}(l_n)$. The Hermitian
matrix $W$ fixes relation between any regular and irregular
Helmholtz solutions. The reactance matrix can be expressed in the
terms of the matrix $W$ as $-K^{L+E/2}W^{-1}K^{L+E/2}$. In the Eq.
($\ref{EqPsi}$) we hold only the leading terms of regular and
irregular solutions. Such asymptotic leads to the boundary
condition on matrix wavefunction
\begin{equation} \label{MatCond}
\left[
\left(\frac{\partial}{\partial r}
\right)^{2L+E} r^{L+E}
\int_{\Omega} A^+({\bf n}) y({\bf r},K{\bf n}_0) d\Omega_{\bf n}
\right]_{r=0}=
\end{equation}
$$
-2^L L! W \left[
\frac{r^{L+E}}{(2L+E)!!} \int_{\Omega} A^+({\bf n})
y({\bf r},K{\bf n}_0) d\Omega_{\bf n}\right]_{r=0}
$$
Breit proved ($\cite{Breit}$, see also
$\cite{Huang,AlGa1988,Balt}$) that
the ZRP and hence matrix ZRP can be also introduced as pseudopotential.

Imposing the boundary condition on the
integral representation of matrix function $y({\bf r},K{\bf n}_0)$
(see Eq. $(\ref{EQy})$) we derive the matrix amplitude
\begin{equation} \label{EQAFA}
F({\bf n},{\bf n}_0)=4\pi\, A({\bf n}) F A^+({\bf n}_0),
\end{equation}
where $F$ is given by
\begin{equation} \label{EQFM_0}
F=-\left(K^{-L} W_0 K^{-L} + iK\right)^{-1},
\end{equation}
and $W_0=(-i)^{L}W i^{L}$ is also hermitian matrix.

Let us consider special cases. \\
$\bullet$ {\it One-channel ZRP\\}
For the case of any state ($l_0=l$) and $W_0=\alpha$
scattering amplitude becomes
$$
F=-\frac{k^{2l}}{\alpha+ik^{2l+1}},
$$
where $\alpha=1/a$ - inverse scattering length.
The expression coincides with scattering amplitude $F$,
which can be calculated for scattering by GZRP $\cite{Balt}$.
Assuming $l=0$ we get scattering amplitude of
isolated ZRP $\cite{DO1975}$.\\
$\bullet$ {\it Two-channel matrix ZRP\\}
Let us consider ground
$s$ ($l_0=0$) and some ($l_1=l$) excited states.
Equation ($\ref{EQFM_0}$) goes into the
inverse matrix amplitude
\begin{equation}\label{EQF0l}
L=
\left(
\begin{array}{cc}
0&0\\0&l
\end{array}
\right),\hspace{8mm}
F^{-1}=-\left(
\begin{array}{cc}
\alpha_0+ik_0& c k_1^{-l} \\
c k_1^{-l} & \alpha_1 k_1^{-2l}+ik_1
\end{array}
\right),
\end{equation}
where $k_0, k_1$ are related by the energy conservation low and
$\alpha_0,\alpha_1,c$ are real parameters. In the case $l=0$ we
obtain one-center matrix amplitude for two $s$ states
$\cite{DO1975}$.

\section{Two matrix ZRP problem} \label{TZRP}
In order to study thoroughly the multiple scattering on a
multi-channel level of approximation we consider two matrix ZRP
problem. Assume two-center matrix potential $U$ is localized on
the points ${\bf R}_1={\bf R},\ {\bf R}_2=-{\bf R}$ and satisfies
the following parity requirement
\begin{equation} \label{EQV}
PUP^{-1}=\Sigma U \Sigma^{-1},
\end{equation}
here $P$ denotes inverse operator $Pf({\bf r})=f(-{\bf r})$, and
$\Sigma=\hbox{diag}(\eta_n)$ is the matrix of parities.

Let us represent $U$ as
$$
U=e^{-i{\bf p}\cdot {\bf R}} U_1 e^{i{\bf p}\cdot {\bf R}} +
e^{i{\bf p}\cdot {\bf R}} U_2 e^{-i{\bf p}\cdot {\bf R}},
$$
where $U_1$, $U_2$ are matrix ZRPs, and $2R$ is a distance between
the centers. This representation leads to
$$
U_2=\Sigma PU_1P^{-1} \Sigma.
$$
Taking into consideration these expressions, we conclude that
\begin{equation}
F_1^{-1}({\bf n},{\bf n}_0)=
F^{-1}({\bf n},{\bf n}_0),
\end{equation}
\begin{equation}
F_2^{-1}({\bf n},{\bf n}_0)=
\Sigma F^{-1}(-{\bf n},-{\bf n}_0) \Sigma.
\end{equation}
In the body-frame, where polar axis $z$ is taken along ${\bf R}$,
matrix $F({\bf n},{\bf n}_0)$ can be represented by
Eq. $(\ref{EQAFA})$. Denoting
\begin{equation} \label{EQCn12}
C_1({\bf n},{\bf n}_0)=(4\pi)^{1/2}\, A({\bf n}) C_1({\bf n}_0),
\hspace{10mm}
C_2({\bf n},{\bf n}_0)=(4\pi)^{1/2}\, A(-{\bf n})
\Sigma C_2({\bf n}_0)
\end{equation}
we obtain for matrixes $C_1({\bf n}_0),C_2({\bf n}_0)$ the equations
\begin{equation} \label{EQC1C2}
\left\{
\begin{array}{l}
F^{-1} C_1({\bf n}_0)-\Sigma H C_2({\bf n}_0)=
(4\pi)^{1/2}\, A^{+}({\bf n}_0) e^{iK{\bf n}_0\cdot{\bf R}},\\
F^{-1} C_2({\bf n}_0)-\Sigma H C_1({\bf n}_0)=
(4\pi)^{1/2}\, \Sigma A^{+}(-{\bf n}_0)^{\displaystyle \mathstrut}
e^{-iK{\bf n}_0\cdot{\bf R}},
\end{array} \right.
\end{equation}
where matrix $H$ (argument $KR$ is omitted) is given by
\begin{equation} \label{integ}
H=4\pi \int_{\Omega}
A^+({\bf n}) H(2{\bf R},{\bf n};K) A(-{\bf n})
d\Omega_{\bf n}.
\end{equation}
In order to calculate matrix
$H$ we can use the following expansion
\begin{equation}
H(2{\bf R},{\bf n};K)=\frac{K}{4\pi}
\sum_{\lambda=0}^{\infty}
i^{\lambda}(2\lambda+1) h_{\lambda}(2KR)\,
P_{\lambda}(R^{-1}{\bf R}\cdot{\bf n}),
\end{equation}
where $\displaystyle h_{\lambda}(x)=x^{\lambda}
\left(-\frac{1}{x}\frac{d}{dx}\right)^{\lambda} \frac{e^{ix}}{x}$
are Riccati-Hankel functions, $P_{\lambda}(x)$ are Legendre
polynomials. The integral $(\ref{integ})$ can be evaluated in
terms of 3-$j$ symbols or Clebsch-Gordan coefficients. Thus,
determination of the matrixes $C_1({\bf n}_0),C_2({\bf n}_0)$ by
using Eqs. $(\ref{EQC1C2})$ reduces to solving a linear system of
algebraic equations. The final result for the matrix amplitude
$F({\bf n},{\bf n}_0)$ becomes
\begin{equation} \label{EQ2cF}
\begin{array}{c}
(2\pi)^{-1} F({\bf n},{\bf n}_0)=\\
(A({\bf n})e^{-iK{\bf n}\cdot{\bf R}} +
\Sigma A(-{\bf n}) e^{iK{\bf n}\cdot{\bf R}})
( F^{-1}\, -\Sigma H )^{-1}
(A^+({\bf n}_0) e^{iK{\bf n}_0\cdot{\bf R}} +
\Sigma A^+(-{\bf n}_0) e^{-iK{\bf n}_0\cdot{\bf R}})+
^{\displaystyle \mathstrut}\\
(A({\bf n})e^{-iK{\bf n}\cdot{\bf R}} -
\Sigma A(-{\bf n}) e^{iK{\bf n}\cdot{\bf R}})
( F^{-1}\, +\Sigma H )^{-1}
(A^+({\bf n}_0) e^{iK{\bf n}_0\cdot{\bf R}} -
\Sigma A^+(-{\bf n}_0) e^{-iK{\bf n}_0\cdot{\bf R}}
)^{\displaystyle \mathstrut}.
\end{array}
\end{equation}
For purposes of illustration, we represent some examples. \\
$\bullet$ {\it One-state level of approximation\\}
Two-center electron-molecular interaction may be
approximated by two ZRPs.
The simplest choice for $\Sigma$-state is two $s$-centers,
i.e. $A({\bf n})=(4\pi)^{-1/2}$.
If one-center inverse scattering
amplitude and $H$ are
$$
F^{-1}=-\alpha-ik,\hspace{10mm} H=\frac{e^{2ikR}}{2R},
$$
then two-center scattering amplitude is given by the expression
$$
F({\bf n},{\bf n}_0)=-2\,
\frac{\cos(k{\bf n}{\cdot}{\bf R})
\cos(k{\bf n}_0{\cdot}{\bf R})}
{\displaystyle \alpha+ik+\frac{e^{2ikR}}{2R}}-2\,
\frac{\sin(k{\bf n}{\cdot}{\bf R})
\sin(k{\bf n}_0{\cdot}{\bf R})}
{\displaystyle \alpha+ik-\frac{e^{2ikR}}{2R}}.
$$
The result coincides completely with scattering amplitude
for two ZRPs $\cite{DO1975}$.\\
$\bullet$ {\it Two-state level of approximation\\} The problem of
electron-impact excitation of the $\Sigma$, $\Pi$, $\Delta$, etc.
molecular states can be considered in two-state level of
approximation. Suppose, for a simplicity sake, molecule have even
ground $\Sigma$-state (i.e. $\eta_0=1$) and any excited state,
which may be $\Sigma$, $\Pi$, $\Delta$, etc. (i.e. $m=0,1,2,$ etc.
and $\eta_1=1$ or $-1$). Let us approximate two-center molecular
interaction by two matrix ZRPs. One-center matrix $F$ is given by
Eq. $(\ref{EQF0l})$, where $l\geq m$. We introduce the following
notation for short
\begin{equation}
\begin{array}{l}
\displaystyle
\theta_0(x)= \alpha_0 +ik_0+ x \frac{e^{2ik_0 R}}{2R},\\
\displaystyle
\theta_1(x)= \alpha_1 +ik_1^{2l+1} + x
\sum_{\lambda=0,2,}^{2l} i^{2l+\lambda}
(2\lambda+1) k_1^{2l+1} h_{\lambda}(2k_1 R)
\int_{\Omega} |Y_{l m}({\bf n})|^2
P_{\lambda}({\bf n}\cdot{\bf e}_z)
d\Omega_{\bf n},
\end{array}
\end{equation}
where ${\bf e}_z$ is unit vector orientated along polar axis $z$.
Thus, formula ($\ref{EQ2cF}$) results in the
elastic scattering amplitude
(i.e. for a transition $0\rightarrow0$)
\begin{equation}
F_{00}({\bf n},{\bf n}_0)=
-2\theta_1(\eta_1)
\frac{\cos(k_0{\bf n}{\cdot}{\bf R})
\cos(k_0{\bf n}_0{\cdot}{\bf R})}
{\theta_0(\eta_0)\theta_1(\eta_1)-c^2}
-2\theta_1(-\eta_1)
\frac{\sin(k_0{\bf n}{\cdot}{\bf R})
\sin(k_0{\bf n}_0{\cdot}{\bf R})}
{\theta_0(-\eta_0)\theta_1(-\eta_1)-c^2}.
\end{equation}
Assuming $\eta_1=-(-1)^l$, we obtain electron-impact
excitation amplitude (for a transition $0\rightarrow 1$)
\begin{equation} \label{EQF10_}
F_{10}({\bf n},{\bf n}_0)=4 \sqrt{\pi} c k_1^{l}
Y_{lm}({\bf n})\left(
\frac{\cos(k_1{\bf n}{\cdot}{\bf R})
\cos(k_0{\bf n}_0{\cdot}{\bf R})}
{\theta_0(\eta_0)\theta_1(\eta_1)-c^2}+
\frac{\sin(k_1{\bf n}{\cdot}{\bf R})
\sin(k_0{\bf n}_0{\cdot}{\bf R})}
{\theta_0(-\eta_0)\theta_1(-\eta_1)-c^2}\right).
\end{equation}
In the case $\eta_1=(-1)^l$ one becomes
\begin{equation} \label{EQF10__}
F_{10}({\bf n},{\bf n}_0)=-4i\sqrt{\pi} c k_1^{l}
Y_{lm}({\bf n})\left(
\frac{\sin(k_1{\bf n}{\cdot}{\bf R})
\cos(k_0{\bf n}_0{\cdot}{\bf R})}
{\theta_0(\eta_0)\theta_1(\eta_1)-c^2}-
\frac{\cos(k_1{\bf n}{\cdot}{\bf R})
\sin(k_0{\bf n}_0{\cdot}{\bf R})}
{\theta_0(-\eta_0)\theta_1(-\eta_1)-c^2}\right).
\end{equation}
According to general theory $\cite{DO1975,D1978}$, a
resonances occur in the vicinity of the poles.
The poles of the matrix amplitudes correspond to the solutions
of the equations
$\theta_0(\pm\eta_0)\theta_1(\pm\eta_1)=c^2.$
Considered as a
function of the spacing on centers $2R$ these energies represent
the adiabatic potential curves of the negative ions or
quasistationary states. In the matrix ZRP model the poles
may reproduce both shape and Feshbach resonances.

The matrix amplitudes are represented in body-fixed frame.
The amplitudes in other frame can be obtained by frame rotation.
The frame-transformation is reduced to the simple substitution
$Y_{lm}({\bf n}) \rightarrow
\sum_{m'=-l}^l D^{l*}_{mm'} Y_{lm'}({\bf n})$ since
the scalar products ${\bf n}_0{\cdot}{\bf R},\,
{\bf n}{\cdot}{\bf R}$ are invariants of a
frame-rotational transformation.
Here $D^l_{mm'}$ denote the rotation matrix elements
(so-called $D$-functions, $\cite{VMK1975}$).

\section{The adiabatic approximation} \label{DICS}
The adiabatic approximation can be used in order that to
incorporate the motion of the nuclei into the theory. Initially
this approximation was applied by Drozdov $\cite{Dr1955}$, Chase
$\cite{Ch1956}$, and Oksyuk $\cite{Ok1965}$. The adiabatic
approximation within the framework of the ZRP model was developed
by Demkov and Ostrovsky $\cite{DO1975}$ and Drukarev and Yurova
$\cite{DYu1977}$. This approximation allows to express the
electron-vibrational transition differential cross section (DCS)
via the electron transition amplitude on the space-fixed matrix
ZRPs:
\begin{equation}
\frac{d\sigma}{d\Omega}(nv\leftarrow 0v_0)=
(4\pi)^{-1}M_n\frac{k_n}{k_0}
\int_{\Omega}
\left|
\left< nv\left| F_{n0}({\bf n},{\bf n}_0,{\bf R})
\right|0v_0\right>
\right|^2 d\Omega_{\bf R},
\end{equation}
where $n,v,0,v_0$ represent electron and vibrational quantum numbers
for final and initial states,
$\langle nv|\cdot|0v_0\rangle$ denotes integral of
the vibrational harmonics, and $M_n$ is the orbital angular
momentum projection degeneracy factor of the final target state $n$.

The integral cross section (ICS)
for electron-vibrational transition is obtained
by the integration over the scattering angle:
\begin{equation}
\sigma(nv\leftarrow 0v_0)=
\int_{\Omega}\frac{d\sigma}{d\Omega}(nv\leftarrow 0v_0)
\, d\Omega_{\bf n}.
\end{equation}
The summation over the vibrational states
(generally - including continuous spectrum) gives the
electron transition ICS
\begin{equation} \label{EQICS}
\sigma(n\leftarrow 0v_0)=
(4\pi)^{-1}M_n \frac{k_n}{k_0}
\left< 0v_0 \left|
\int_{\Omega} \int_{\Omega} |F_{n0}({\bf n},{\bf n}_0,{\bf R})
|^2\, d{\Omega}_{\bf n}d{\Omega}_{{\bf n}_0}\right| 0v_0 \right>.
\end{equation}
In these equations the integrals over angles can be reduced to
the spherical harmonics sums, which are most suitable for
numerical calculation. \\
$\bullet$ {\it The electron-vibration transition DCS for
two-channel problem\\} It is convenient to represent the
amplitudes Eqs. $(\ref{EQF10_},\ref{EQF10__})$ in coordinate
frame, where the polar axis is taken along vector ${\bf n}$. As it
proved above, the rotational transformation is reduced to the
substitution $Y_{lm}({\bf n})\rightarrow Y_{lm}({\bf R})$.
Expanding the amplitudes in the series over spherical harmonics we
obtain the expression
$$
\frac{d\sigma}{d\Omega}(nv\leftarrow 0v_0)=
(4\pi)^2\, M_n\frac{k_n}{k_0} \sum_{\lambda=0}^{\infty}
\sum_{\lambda'=0}^{\infty}
\sum_{\mu=-\min(\lambda,\lambda')}^{\min(\lambda,\lambda')}
Q_{\lambda\mu}Q^*_{\lambda'\mu}
\int_{\Omega} |Y_{l_n m_n}({\bf R})|^2 Y^*_{\lambda\mu}({\bf R})
Y_{\lambda'\mu}({\bf R}) \, d\Omega_{\bf R},
$$
where integral of the spherical harmonics can be evaluated in terms of
the Clebsch-Gordan coefficients, and $Q_{\lambda\mu}$ are given by
$$
\begin{array}{c}
Q_{\lambda\mu}=
\displaystyle \frac{i^{\lambda}(-1)^{l_n}\eta_n+(-i)^{\lambda}}{2}
\times\\
\displaystyle \left(
Y_{\lambda\mu}(k_n{\bf n}+k_0{\bf n}_0)
\left< nv \left| \left(Z_n^{(+)}-Z_n^{(-)}\right)
j_{\lambda}(|k_n{\bf n}+k_0{\bf n}_0|R)
\right|0v_0\right>^{\mathstrut} + \right. \\ \left.
\displaystyle Y_{\lambda\mu}(k_n{\bf n}-k_0{\bf n}_0)
\left< nv \left|
\left( Z_n^{(+)}+Z_n^{(-)}\right)
j_{\lambda}(|k_n{\bf n}-k_0{\bf n}_0|R)\right|
0v_0 \right>^{\mathstrut} \right),
\end{array}
$$
where
$$
Z_{n}^{(+)}=\frac{\left(-\theta_1(\eta_1),\ ck_1^{l_1}\right)}
{\theta_0(\eta_0)\theta_1(\eta_1)-c^2},
\hspace{8mm}
Z_{n}^{(-)}=\frac{\left(-\theta_1(-\eta_1),\ ck_1^{l_1}\right)}
{\theta_0(-\eta_0)\theta_1(-\eta_1)-c^2}.
$$
$\bullet$ {\it The electron transition ICS for two-channel
problem\\} Substitution Eqs. $(\ref{EQF10_},\ref{EQF10__})$ into
Eq. $(\ref{EQICS})$ yields the following ICSs
$$
\begin{array}{r}
\displaystyle \sigma(n\leftarrow 0v_0)= 4\pi M_n \frac{k_n}{k_0}\,
\left< 0v_0 \left| |Z_n^{(+)}|^2 \left(1+\eta_0
\frac{\sin(2k_0R)}{2k_0R}\right) \left(1+(-1)^{l_n} \eta_n B_{l_n
m_n}(2k_n R) \right)+ \right. \right. \\  \displaystyle \left.
\left. |Z_n^{(-)}|^2 \left(1-\eta_0
\frac{\sin(2k_0R)}{2k_0R}\right) \left(1-(-1)^{l_n} \eta_n B_{l_n
m_n}(2k_n R) \right) \right| 0v_0\right>^{\mathstrut}.
\end{array}
$$
$$
\begin{array}{c}
B_{lm}(x)= \displaystyle \sum_{\lambda=0,2,}^{2l}
i^{\lambda}(2\lambda+1) j_{\lambda}(x) \int_{\Omega} |Y_{lm}({\bf
e})|^2 P_{\lambda}({\bf e}\cdot{\bf e}_z) d\Omega_{\bf e}.
\end{array}
$$

\section{Applications and discussion} \label{Apl}

\begin{figure}
\centering
\includegraphics[scale=0.9]{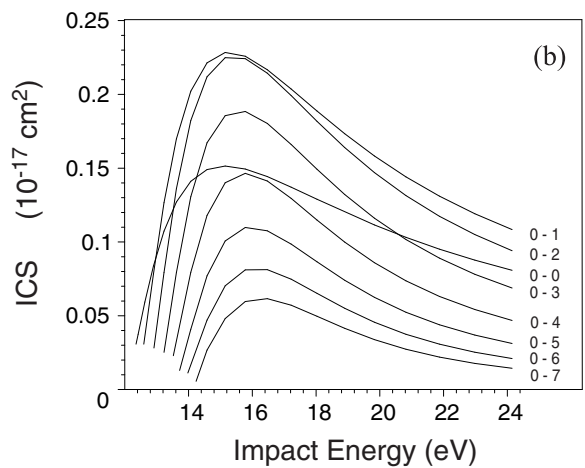}
\parbox[t]{0.7\textwidth}{
\caption{Total cross sections for excitation of $v=0,1,..,7$
vibrational levels of the $a ^3\Sigma_g$ state of $\hbox{H}_2$.}
\label{f:ics_H2_a3S_g_v}}
\end{figure}

The integral cross section for electron-vibrational excitation of
$a\, ^3\Sigma_g^{+}$ state of $\hbox{H}_2$ is plotted in fig.
$\ref{f:ics_H2_a3S_g_v}$ for a number of values of $\alpha_0,
\alpha_1, c, R$ (see $\cite{LY}$), which are regarded as constant
in the range of interest.

\section{Conclusion}
Among the most important aspects of the paper are the development
of the matrix ZRP theory, the calculation of the differential and
integral cross sections for the electron-vibrational transitions,
and investigation of the matrix ZRP possibilities. The
nonoverlapping condition essentially simplifies a scattering
problem solving. Thus, the determination of the multi-center
matrix amplitude reduces to solving of the system (see Eqs.
$(\ref{EQC}),(\ref{EQFmc})$) of integral equations. The number of
scattering centers defines the number of equations in the system.
\section{Acknowledgements}
   We acknowledge consultations  of
V. Ostrovsky and I. Yurova and discussions with J. Sienkiewicz and
M. Zubek.


\begin{thebibliography}{99}
\bibitem{Fermi}
         Fermi, E., 1936. Ric. Sci. 7, 13-52.
\bibitem{Breit}
         Breit, G., 1947. Phys. Rev. 71, 215-31.
\bibitem{DO1975}
         Demkov, Yu., N., Ostrovsky, V., N., 1988.
         Zero-Range Potentials and their Applications in Atomic
         Physics, Plenum, New York.
\bibitem{DYu1977}
         Drukarev, G., F., Yurova, I., Yu., 1977. J. Phys. B: At. Mol.
         Phys. 10, 3551-8
\bibitem{D1978}
         Drukarev, G., F., 1978. Adv. Quantum Chem. 11, 251
\bibitem{AlGa1988}
         Albeverio, S., Gesztesy, F., H{\o}egh-Krohn, R., Holden, H., 1988. Solvable
         Models in Quantum Mechanics, Springer-Verlag, New York.
\bibitem{Huang}
         Huang, K., Yang, C., N., 1957. Phys. Rev. 105, 767
\bibitem{Prunele}
         De Prunel$\acute{\hbox{e}}$, E.,
         1997, J. Phys. A: Math. Gen. 30, 7831-7848
 \bibitem{BW}
         Blatt, J., M., Weisskopf, V., F., 1952.
         Theoretical Nuclear Physics Wiley, New York, Section 2.3.C
\bibitem{Balt}
         Baltenkov, A., S., 2000. Phys. Lett. A, 286, 92-99.
\bibitem{DO1970}
         Demkov, Yu., N., Ostrovsky, V., N., 1970.
         Sov. Phys.-JEPT, 32, 959-63
\bibitem{LY}
         Leble, S., B., Yalunin, S., preprint quant-ph/0205110
\bibitem{Dr1955}
         Drozdov, S., I., 1955, Sov. Phys.-JETP, 1, 591-2
\bibitem{Ch1956}
         Chase, D., M., 1956. Phys. Rev., 104, 838-42
\bibitem{Ok1965}
         Oksyuk, Yu., D., 1965. Sov. Phys.-JETP, 49, 1261-73
\bibitem{La1980}
         Lane N., F., 1980. Rev. Mod. Phys., 52, 29-119
\bibitem{VMK1975}
         Varshalovich, D., A., Moskalev, A., N., Khersonskii, V.,
         K., 1975. Quantum Theory of Angular Momentum, Nauka, Leningrad
         (in Russian)
\end{thebibliography}
\end{document}